\documentclass[fleqn,twoside]{article}
\usepackage{espcrc2,epsfig}

\usepackage{graphicx}
\usepackage[figuresright]{rotating}

\newcommand{\AmS}{{\protect\the\textfont2
  A\kern-.1667em\lower.5ex\hbox{M}\kern-.125emS}}

\def\Journal#1#2#3#4{{#1} {#2}:{#3} (#4)}

\def\PRL{\em Phys. Rev. Lett.}

\def\ARNP{\em Annu. Rev. Nucl. Part. Sci.}

\def\ie{\mbox{{\it i.e.}}}

\newcommand{\inmath}[1] {\ifmmode#1\else$#1$\fi}
\newcommand{\definmath}[2] {\def#1{\ifmmode#2\else$#2$\fi}}

\newcommand{\Mw}{\mbox{$M_W$}}

\newcommand{\Mt}{\mbox{$M_t$}}
\newcommand{\MT}{\mbox{$M_T$}}

\newcommand{\roots}{\mbox{$\sqrt{s}$}}

\title{Electroweak Prospects for Tevatron RunII}

\author{D. Glenzinski \small (on behalf of the CDF and D0 Collaborations)
  \address[FNAL]{Fermi National Accelerator Laboratory, P.O. Box 500,
     Batavia, Il. 60501-0500 U.S.A
  }
}
       
\begin{document}

\begin{abstract}
The prospects of precision electroweak measurements from CDF and D0 using
RunII data is reviewed.
\vspace{1pc}
\end{abstract}

\maketitle

% --- INTRODUCTION

\section{Introduction}

In RunI each experiment collected about 
$100\:\mathrm{pb}^{-1}$ of data.  During RunIIa, each experiment is expected 
to collect about $2\:\mathrm{fb}^{-1}$ of data.  The center-of-mass 
energy for RunII, $\roots=2.0$~TeV, is a bit larger than the
$1.8$~TeV of RunI and results in an increase of about 10\% (35\%) in the
production cross-sections for $W$ and $Z$ ($t\overline{t}$) events.  Additional
gains in the event yield are expected due to improvements in 
the detector acceptance and performance.  Taken together, the RunIIa upgrades
are expected to yield 2300k (800) $W$ ($t\overline{t}$) events per experiement,
including the effects of event selection and triggering, which can be compared 
to the RunI yields of 77k (20) events.  With the RunI data-set, CDF and D0 
produced a breadth of electroweak results and obtained the world's only sample 
of top quarks. While the RunII electroweak physics program is very similar,
the RunII upgrade improvements should yield many precision results.

The Tevatron began delivering steady data in about June, 2001.  The first
six months of data taking was ``commissioning dominated'' for CDF and D0.
Starting around January, 2002, the experiments were largely commissioned and
began taking ``analysis quality'' data.  The physics results reported at this
conference are based on about $10-20\:pb^{-1}$ (depending on the data-set) 
per experiment.  Thus, the presently available event samples are smaller than 
those available in RunI.  At this early stage of RunII, it is interesting to 
compare the present detector performance to that assumed when making the RunII
physics projections.  

In the following sections I discuss some RunII projections for a few 
electroweak measurements of particular importance, namely the precision 
determinations of the W-boson mass, \Mw, and the top-quark mass, \Mt.

%%%\begin{table*}[htb]
%%%\caption{Approximate event yields per experiment, including the effects of 
%%%selection algorithms and triggers.}
%%%\label{tab:yields}
%%%\newcommand{\m}{\hphantom{$-$}}
%%%\newcommand{\cc}[1]{\multicolumn{1}{c}{#1}}
%%%\renewcommand{\tabcolsep}{2pc}    % --- enlarge column spacing
%%%\renewcommand{\arraystretch}{1.2} % --- enlarge line spacing
%%%\begin{tabular}{@{}ccc}\hline
%%%Sample &RunI &RunII \\ \hline
%%%$W\rightarrow\ell\nu$ &77k &2300k \\
%%%$Z\rightarrow\ell\ell$   &10k &202k  \\
%%%$WV$ ($W\rightarrow\ell\nu$, $V=W,\gamma,Z$)
%%%                         &90  &1800  \\
%%%$ZV$ ($Z\rightarrow\ell\ell$, $V=W,\gamma,Z$)
%%%                         &30  &500   \\
%%%$t\overline{t}$ (mass sample,$\geq1$ B-tag)
%%%                         &20  &800   \\ \hline
%%%\end{tabular}\\ [2pt]
%%%In the Table, $\ell\equiv\mathrm{e},\mu$.
%%%\end{table*}

% --- Mw

\section{Precision Measurement of the W-Boson Mass}

In RunI, CDF and D0 each measured the W-Boson mass with an uncertainty of
about $80$~MeV, which together yield a combined Tevatron result of 
$80.456\pm0.059$~GeV~\cite{tevmw}.  For comparison, the (preliminary) LEP 
combined result is $80.447\pm0.042$~GeV~\cite{lepmw}.

At the Tevatron the W-bosons are produced by hard collisions between the
constituent (anti-)quarks of the proton and anti-proton.  Thus, the 
center-of-mass energy for this hard collision cannot be known {\it{ a priori}}
event-by-event.  As a consequence of this, only the constraints
in the transverse plane remain.  Due to overwhelming QCD backgrounds, only
leptonically decaying Ws are used.  The W-boson mass can be extracted from a 
fit to the transverse mass, 
$M_{T}= \sqrt{(E^{\ell}_{T}+E^{\nu}_{T})^{2}-
(\overline{P}^{\ell}_{T}+\overline{P}^{\nu}_{T})^{2}}$.  It is obvious that
a detailed understanding of the energy and momentum scales and resolutions
is fundamental to this measurement.  In order to accurately estimate the
neutrino momentum, it is additionally important to have a detailed
understanding of the underlying event and recoil distributions.  Finally,
since the shape of the \MT\ distribution is affected by the 
parton-distribution-functions (PDFs), there is a model-dependent systematic
uncertainty associated with this method.  For a detailed discussion of these
measurements, see references~\cite{tevmw} and~\cite{mwreview}.

\begin{table*}[htb]
\caption{Uncertainties in the CDF/D0 combined \Mw\ measurement.
  The contribution of each systematic source is approximate.}
\label{tab:tevmw}
\newcommand{\m}{\hphantom{$-$}}
\newcommand{\cc}[1]{\multicolumn{1}{c}{#1}}
\renewcommand{\tabcolsep}{2pc}    % --- enlarge column spacing
\renewcommand{\arraystretch}{1.2} % --- enlarge line spacing
\begin{tabular}{@{}rc}\hline
 \cc{Source}       & $\Delta\Mw$ (MeV)  \\ \hline
 Statistical: & $\pm 40$           \\ \hline
 Systematic   &                    \\
 scale:       & $\pm 40^{\dagger}$ \\
 recoil:      & $\pm 20^{\dagger}$ \\
 modeling:    & $\pm 15^{*}$       \\
 other:       & $\pm 15^{\dagger}$ \\
 total:       & $\pm 43\:\:$       \\ \hline
\end{tabular} \\ [2pt]
$\dagger$ dominated by statistics of control sample. \\
$*$ correlated between experiments.
\end{table*}

The sources of uncertainty in the combined Tevatron \Mw\ determination
are given in Table~\ref{tab:tevmw}.   The ``other'' category includes 
uncertainties from background shape and normalization and residual fit biases.
As is evidenced in the Table, even the Tevatron combined measurement is
dominated by uncertainties which are expected to scale with the statistics
of the relevant data-sets.  The only exceptions to this are the ``modeling''
uncertainties, dominated by contributions from PDF 
uncertainties, but also including contributions from higher-order radiative 
corrections.  As these are the only sources of correlated uncertainties between
the experiments, they may well come to limit the ultimate precision with which
\Mw\ can be determined at the Tevatron.  Assuming no improvement in these
modeling uncertainties (\ie\ a conservative assumption), it is rather straight-
forward to appropriately scale the other uncertainties to arrive at a 
projection of $\Delta\Mw(2\:\mathrm{fb}^{-1})=\pm30\:\mathrm{MeV}$ per 
experiment and a Tevatron combined uncertainty of about $25\:\mathrm{MeV}$.
The only important caveat in this projection is the assumed resolution on
$P_{T}^{\nu}$, which degrades with the mean number of additional interactions
per event.  This functional dependence was studied using RunI data.  Although
the RunIIa instantaneous luminosity increases, the mean number of additional 
interactions per event is comparable to that of RunI because the Tevatron is 
running with more bunches.  Thus, for the RunIIa projection, this assumption 
is on fairly solid footing.  Finally, depending on the evolution of the 
combined LEP uncertainty, and assuming the LEP and Tevatron measurements are 
completely uncorrelated, this gives an expected world average of 
$\Delta\Mw(\mathrm{LEPII+TeVIIa})=\pm15-20\:\mathrm{MeV}$.

% --- Mtop

\section{Precision Measurement of the Top-Quark Mass}

In RunI, CDF and D0 each measured the top-quark mass with an uncertainty of
about $7$~GeV, which together yield a combined Tevatron result of 
$174.3\pm5.1$~GeV~\cite{tevmt}.

At the Tevatron, top quarks are predominantly pair-produced, with each
top quark predominantly decaying to a $W$ and a b quark.  The final-state
topology is determined by the decay of the two $W$s, with the ``di-lepton'',
``lepton plus jets'' and ``fully hadronic'' final states corresponding to 
both, one, or neither of the $W$s decaying leptonically, respectively.  While
the di-lepton final state is the most pure, it's branching ratio is smallest
(owing to the $BR(W\rightarrow\ell\nu)^2$ factor) and it's kinematics are
under constrained (owing to the two neutrinos).  On the other hand, the fully
hadronic final state suffers from a large QCD background.  Consequently, the
most significant channel for the determination of \Mt\ is the
lepton-plus-jets channel.  The dominant background contributions to
this channel are from $W$+$\geq$4~jet events, which can be suppressed by
requiring that $\geq1$~jet in the event is identified as a b-quark jet
(``B-tagged'').
The top-quark mass for each candidate event is determined from a kinematic
fit which employs momentum constraints and requires that the two $W$ candidates
have a mass consistent with the world average \Mw, and that the two top quarks
have the same mass.   In order to perform the fit, jet-parton assignments must
be made, thus giving rise to a combinatoric background, which greatly
degrades the resolution of the kinematic fit. For lepton-plus-jet
events with 0, 1 or 2 B-tagged jets, there are 12, 6 or 2 possible jet-parton
combinations.  Thus, the single most important factor in improving the \Mt\
determination for RunII, is the expected improvement in the B-tagging 
performance, which should yield a more efficient and pure event selection,
and should reduce the combinatoric background, effectively enhancing the
per event \Mt\ sensitivity.  For a detailed discussion of these
measurements, see references~\cite{tevmt} and~\cite{mtreview}.

\begin{table*}[htb]
\caption{``Typical'' per experiment uncertainties in the RunI \Mt\ 
  measurement.}
\label{tab:mtop}
\newcommand{\m}{\hphantom{$-$}}
\newcommand{\cc}[1]{\multicolumn{1}{c}{#1}}
\renewcommand{\tabcolsep}{2pc}    % --- enlarge column spacing
\renewcommand{\arraystretch}{1.2} % --- enlarge line spacing
\begin{tabular}{@{}rc}\hline
 \cc{Source}       & $\Delta\Mt$ (GeV)  \\ \hline
 Statistical: & $\pm 5\:\:$       \\ \hline
 Systematic   &                   \\
 scale:       & $\pm 4\:\:$       \\
 modeling:    & $\pm 2^{*}$       \\
 other:       & $\pm 2\:\:$       \\
 total:       & $\pm 5\:\:$       \\ \hline
\end{tabular} \\ [2pt]
$*$ correlated between experiments.
\end{table*}

The sources of uncertainty in the ``typical'' RunI \Mt\ determination
are given in Table~\ref{tab:mtop}.   The ``other'' category includes 
uncertainties from background shape and normalization and residual fit biases.
The ``modeling'' uncertainties are dominated by contributions from 
hadronization and fragmentation modeling, and modeling of final state
gluon radiation and are the only source of correlated uncertainty between
the experiments.  The dominant systematic uncertainty is due to uncertainties
in the jet energy scale and associated corrections, which, in RunI, were 
determined from low statistics control samples. For $2\:\mathrm{fb}^{-1}$ of
RunII data, the statistical uncertainty is expected to be $<1\:\mathrm{GeV}$
per experiment and the measurements are expected to be systematic limited.
The RunIIa projections assume that the total systematic uncertainties can
be reduced to the $2-3\:\mathrm{GeV}$ per experiment.  Reducing the systematic
uncertainty to that level will require the use of special control samples
($Z+$~jets, $Z\rightarrow b\overline{b}$, and $W\rightarrow q\overline{q}$) 
which, in general, were too small to be of use in RunI.  Thus, the RunII
projections have been based on detailed Monte Carlo simulations of these
data-sets.  The efficiency and purity with which many of these controls samples
are collected are contingent upon the performance of the silicon vertex
detectors.  There have not been, to the author's knowledge, any detailed
study to estimate the projected Tevatron combined \Mt\ uncertainty after
$2\:\mathrm{fb}^{-1}$ of data.  Obviously the projection is strongly dependent
upon the assumed evolution of the modeling uncertainty.  The most conservative
projection would assume no improvement, so that each experiment would have a
measurement uncertainty dominated by these modeling uncertainties.  In that
scenario, the combination would yield very little improvement over the 
$2-3\:\mathrm{GeV}$ uncertainty per experiment.

% --- DETECTORS

\section{Initial Detector Performance}

The CDF and D0 detector upgrades have been described many times and the
details are available in references~\cite{cdftdr}~\cite{dzerotdr}.  It is
principally important to note that the RunII electroweak projections assume 
i) the energy and momentum resolutions are no worse than those of RunI, 
ii) B-jet
and lepton\footnote{Through the whole of this note, it should be understood 
that ``lepton'' means an electron or a muon, unless otherwise stated.} 
identification are extended to the $\left|\eta\right| >1$ forward regions 
\footnote{The co-ordinate system has the z-axis parallel to 
the beam-axis and pointing in the proton flight direction, the x-axis 
orthogonal to the z-axis and pointing
to the center of the Tevatron, and the y-axis defined to yield a right-handed
co-ordinate system;  the angles $\theta$ and $\phi$ are the traditionally 
defined polar and azimuth spherical co-ordinates, respectively, and
the pseudo-rapidity, $\eta$, is defined as $\eta=-\ln(\tan(\theta/2))$.}, 
and iii) the trigger performance
allows efficient collection of the relevant samples up to instantaneous
luminosities of about $2\times10^{32}\:\mathrm{cm}^{-2}\mathrm{sec}^{-1}$.
The performance of the relevant detector components is briefly discussed here.

Both CDF and D0 upgrades include silicon micro-strip detectors at inner
radii, surrounded by large volume tracking chambers, all inside a magnetic
field.   Both experiements have finished initial alignments of their silicon 
detectors and are measuring high signal-to-noise ratios ($\geq12$) with the 
expected intrinsic resolution and high hit efficiency ($\geq98\%$).  This 
bodes well for the B-tag performance of the two experiments.  Both experiments
have also collected large statistics samples of $J/\Psi\rightarrow\mu\mu$
events, which are used to perform a variety of systematic studies to
limit residual mis-alignments, and determine energy-loss and B-field 
corrections.  Although still at an early stage,
the present $P_{T}$ resolution for CDF's COT is better than 
$\sigma_{P_{T}}/P_{T}^{2}<0.13\%\:\mathrm{GeV}^{-1}$, comparable to the
$0.10\%$ design goal~\cite{nahn};  this is expected to improve as 
the alignment matures.
Similarly, D0 expects the CFT to meet design goals once alignment and
calibrations are finalized.  

Since both experiments left their calorimetry largely unchanged relative to
RunI (CDF replaced their forward calorimetry), the resolution should be 
well understood.  The observed width of the invariant mass spectrum in
$Z\rightarrow e^{-}e^{+}$ events can be used to estimate this resolution.
Including all major corrections for both the forward and central 
calorimeters, CDF observes a width within $5\%$ of that 
expected~\cite{erbacher}.  At the 
time of the conference, D0 had not yet included the full set of corrections 
and, consequently, were observing a width $30\%$ larger than expected; this is
roughly consistent with the contribution expected from the excluded 
corrections.  

Despite some initial problems, both experiments have 
demonstrated their ability to efficiently trigger on events of interest.  CDF
has measured a Level 1 tracking efficiency $>95\%$ (important for triggering
on high-momentum leptons) and D0 showed similarly high effiencies for 
high-momentum electrons.  CDF has also collected large samples enriched in
heavy-flavor (c- and b-quark jets) by use of a displaced track trigger at
Level 2 (SVT) - the first such trigger at a hadron collider~\cite{cerri}.  
This is an important first step in accumulating the 
$Z\rightarrow b\overline{b}$ control sample mentioned above.  

Although it is still a bit early to draw any definitive
conclusions, both the CDF and D0 detectors look to be on track to meet their
design goals and fulfill the RunIIa Electroweak physics projections.

%
%%%\begin{figure}[tbhp]
%%%  \begin{center}
%%%    \epsfxsize = 3in
%%%    \epsffile{jpsi_mass.eps}
%%%  \end{center}
%%%  \caption{CDF J/$\Psi$ cadidate events.}
%%%  \label{fig:jpsi}
%%%\end{figure}

% --- END DOCUMENT
\end{document}